\documentclass[11pt,onecolumn,amssymb,nofootinbib]{revtex4}
\usepackage{amsmath, amsthm, amscd, amssymb}
\usepackage{bm}
\usepackage{bbm}
\usepackage{color}

\begin{document}

\title{Canonical Field Anticommutators in the Extended Gauged Rarita-Schwinger Theory  }

\author{Stephen L. Adler}
\email{adler@ias.edu}
\affiliation{Institute for Advanced Study,
Einstein Drive, Princeton, NJ 08540, USA}

\author{Marc Henneaux}\email{henneaux@ulb.ac.be}
\affiliation{Universit\'e Libre de Bruxelles and International Solvay Institutes, ULB-Campus Plaine
 CP231, 1050 Brussels, Belgium }

\author {Pablo Pais}
\email{pais@cecs.cl}
\affiliation{{Centro de Estudios Cient\'{\i}ficos (CECs), Av.~Arturo Prat~514, Valdivia, Chile }\\
Universit\'e Libre de Bruxelles and International Solvay Institutes, ULB-Campus Plaine
 CP231, 1050 Brussels, Belgium }

\begin{abstract}
We reexamine canonical quantization of the gauged Rarita-Schwinger theory using the extended theory, incorporating a  dimension $\frac{1}{2}$ auxiliary  spin-$\frac{1}{2}$ field
$\Lambda$, in which there is an exact off-shell gauge invariance.  In  $\Lambda=0$ gauge, which reduces to the original unextended theory,
our results agree with those found by Johnson and Sudarshan, and later verified by Velo and Zwanziger, which give a canonical Rarita-Schwinger field Dirac bracket that is singular for small gauge fields.   In gauge covariant radiation gauge, the Dirac bracket  of the Rarita-Schwinger fields is nonsingular, but does not correspond to a  positive semi-definite anticommutator, and the Dirac bracket of the auxiliary fields has a singularity of the same form as found in the unextended theory.  These results indicate that gauged Rarita-Schwinger theory is somewhat pathological, and cannot be canonically quantized within a conventional positive semi-definite metric Hilbert space.  We leave open the questions of whether consistent quantizations can be achieved by using an indefinite metric Hilbert space,  by path integral methods, or by appropriate couplings to conventional dimension $\frac{3}{2}$ spin-$\frac{1}{2}$  fields.
\end{abstract}


\maketitle

\section{Introduction}
The conventional approach to grand unification of the strong and electroweak forces assumes that gauge anomalies are to be cancelled among
spin-$\frac{1}{2}$ fermion fields.  So far, no definitive solution to the grand unification problem has been achieved within this framework, raising
the question of whether the rules for constructing unification models should be broadened, and in particular whether there may be more general mechanisms for gauge anomaly cancelation.  In 1985 Marcus \cite{marcus} noted that there are $SU(8)$ representations that cancel anomalies among spin-$\frac{1}{2}$
and spin-$\frac{3}{2}$ fields, assuming that the gauge anomalies for spin-$\frac{3}{2}$ fields are a factor of 3 times the corresponding anomalies
for spin-$\frac{1}{2}$ fields,  and Adler \cite{adler1} recently constructed a concrete model incorporating this observation.  However, anomaly
cancellation using spin-$\frac{3}{2}$ raises again the old question, first explored by Johnson and Sudarshan \cite{johnson}, and by Velo and Zwanziger \cite{velo}, of whether gauged Rarita-Schwinger field theory is consistent in the first place,  either as a classical theory or as a
quantized theory.

With these motivations, the consistency of gauged Rarita-Scwhinger theory has been recently investigated by Adler \cite{adler2}, \cite{adler3}.\footnote{For another recent study, see Dengiz \cite{dengiz}.} In
\cite{adler2} he showed that the problem of superluminal propag{a}tion, found by Velo and Zwanziger in Rarita-Schwinger theory with kinematic mass terms
that do not arise through spontaneous symmetry breaking, is absent in the massless theory.  In \cite{adler3}, he showed that imposing a fermionic analog of the covariant radiation gauge condition leads to a Dirac bracket for the Rarita-Schwinger fields that corresponds on quantization to a
positive semi-definite anticommutator.  However, the assumption that such a gauge condition can be imposed is {\it ad hoc}, and subject to question, because the gauged Rarita-Schwinger theory admits a fermionic gauge
invariance only on-shell, and not off-shell.  Our purpose in the present paper is to re-examine the Dirac bracket calculation of \cite{adler3},
using the extended Rarita-Schwinger theory formulated in \cite{adler2}, in which through adding a dimension $\frac{1}{2}$ spin-$\frac{1}{2}$ auxiliary
field $\Lambda$, an exact off-shell fermionic gauge invariance is achieved. In $\Lambda=0$ gauge, the extended theory reproduces the original results of
\cite{johnson} and \cite{velo}. In covariant radiation gauge, we find that although certain calculations of \cite{adler3} carry over into the extended theory, extra terms are present which spoil positivity of the anticommutator that corresponds to the Rarita-Schwinger field Dirac
bracket.  We also find in radiation gauge that the auxiliary field Dirac bracket has a singularity for small fields that corresponds to the singular
behavior found in \cite{johnson} and \cite{velo}.

The issues discussed in this paper have not been dealt with previously in the literature.
There is extensive literature
showing that spin-$\frac{3}{2}$ fields are consistent within the context of supergravity, where the number of fermionic propagating degrees of freedom is not increased by the interaction; see for example \cite{gris1} and \cite{gris2}.  However, there is no supergravity theory incorporating general $SU(N)$, and in particular $SU(8)$, gauge fields.   The $N$-extended supergravity theories
incorporate a $SO(N)$ vector multiplet for $N=2,...,8$. The maximum number of vector fields in the spin-$\frac{3}{2}$ supermultiplet that are available for ``gauging" is limited to $28$, which occurs for maximal ($N=8$) supergravity. Moreover, in the paper of Freedman and Das constructing a gauged $SO(3)$ supergravity \cite{das}, the authors  explicitly state that their ``perturbative calculations do not directly address previous difficulties'' found in the earlier literature
by Johnson and Sudarshan \cite{johnson}  and by Velo and Zwanziger \cite{velo}.  Similar comments on the absence of a proof that ``helicity 3/2 fields can interact consistently only if they belong to the graviton supermultiplet" were expressed by Strathdee \cite{strath1}. On the other hand, there are a number of calculations in
the literature of the spin-$\frac{3}{2}$ non-Abelian gauge anomaly \cite{nielsen1},\cite{duff0},\cite {witten},\cite{nielsen2} and gravitational anomaly \cite{duff0},\cite{witten},\cite{duff1},\cite{duff2},\cite{nielsen3}  for a general non-Abelian gauge group.  If the spin-$\frac{3}{2}$ non-Abelian gauge anomaly is calculable for a general non-Abelian gauge group, then by implication the quantization of a spin-$\frac{3}{2}$ field with
general non-Abelian gauging should be consistent for at least some version of the spin-$\frac{3}{2}$ theory, but this has never been demonstrated in the literature.
Thus a study of the consistency of gauged Rarita-Schwinger fields, as undertaken in this paper, is warranted.

This paper is organized a follows.  In Sec. { \ref{section_free_RS} } we review the ungauged Rarita-Schwinger theory, which has an off-shell fermionic gauge invariance,
and count the degrees of freedom. In Sec. { \ref{section_gauged_RS} } we generalize to the gauged Rarita-Schwinger theory, in which the fermionic gauge invariance is
only on-shell, and show that there are additional degrees of freedom.  In Sec.  { \ref{section_extended_RS} }  we present the extended Rarita-Schwinger theory introduced
in \cite{adler3}, which has a full  off-shell fermionic gauge invariance.  The additional degrees of freedom noted in Sec. { \ref{section_gauged_RS} } are now accounted for
by the auxiliary field $\Lambda$, and  the second class constraints found in Sec. { \ref{section_gauged_RS} } have now become first class by virtue of contributions from
the auxiliary field.  In Sec. { \ref{section_Lambda_0}  }  we impose $\Lambda=0$ gauge, and show that the formalism reproduces the results of \cite{johnson} and
\cite{velo} for the Rarita-Schwinger field Dirac bracket.  In Sec.  { \ref{section_radiation_gauge} }, we impose an analog of radiation gauge natural to the case when the
auxiliary field is non-zero, and compute the Rarita-Schwinger field and auxiliary field Dirac brackets.  In Sec. { \ref{section_failure_positivity} }, we show that the corresponding
anticommutators for the Rarita-Schwinger and auxiliary fields are not positive semi-definite, and in fact, the gauge field averaged anticommutator
for the auxiliary field is negative semi-definite. In Sec. { \ref{section_path_integral} } we formulate path integral quantization in covariant radiation gauge, and in Sec. { \ref{section_discussion} } we state brief conclusions.  Some useful identities from \cite{adler2} that are used in
the calculations of this paper  are summarized in Appendix \ref{appendix}.

\section{The free Rarita-Schwinger theory}
\label{section_free_RS}

We start from the action for the non-interacting classical Rarita-Schwinger field, given in left chiral two-component spinor form \cite{adler2} by

\begin{equation}\label{leftactionf}
S=\frac{1}{2}\int d^4x  [-\Psi_{0}^{\dagger} \vec \sigma \cdot \vec \nabla\times \vec {\Psi}
+\vec {\Psi}^{\dagger} \cdot \vec \sigma \times \vec \nabla \Psi_{0}
+\vec{\Psi} ^{\dagger} \cdot \vec \nabla \times \vec \Psi - \vec{\Psi}^{\dagger} \cdot \vec \sigma \times \partial_{0} \vec{\Psi}]~~~.
\end{equation}
This action is invariant under the fermionic gauge transformation
\begin{align}\label{gaugef}
\vec \Psi \to& \vec \Psi + \vec \nabla \epsilon~~~,\cr
\Psi_0 \to& \Psi_0+ \partial_0 \epsilon~~~,\cr
\end{align}
with $\epsilon$ a fermionic gauge parameter.  This gauge invariance holds off-shell, that is without using the Euler-Lagrange
equations following from varying the action of Eq. \eqref{leftactionf}.

Varying with respect to $\vec{\Psi}^{\dagger}$ we get the Euler-Lagrange equation for $\Psi$,
\begin{equation}\label{eulerf}
0=\vec \sigma \times \vec \nabla \Psi_{0} + \vec \nabla \times \vec{\Psi}-\vec \sigma \times \partial_{0}\vec {\Psi}~~~,
\end{equation}
while varying with respect to $\Psi_{0}^{\dagger}$ we get the constraint
\begin{equation}\label{Kf}
0=K \equiv \frac{1}{2}\vec \sigma \cdot \vec \nabla \times \vec{\Psi}~~~.
\end{equation}
Varying with respect to $\Psi_{0}$ gives, after integrating  Eq. \eqref{leftactionf} by parts, the
adjoint constraint
\begin{equation}\label{adjKf}
0=K^{\dagger} \equiv -\frac{1}{2} {\vec \Psi}^{\dagger}\cdot (\overleftarrow \nabla \times \vec \sigma)
\end{equation}

The action of Eq.\eqref{leftactionf} can now be rewritten in a form that exhibits the Hamiltonian $H$,
\begin{align}\label{lagrang}
S=&\int dt L ~~~,\cr
L=&\int d^3x({\rm time~derivatives} + {\rm constraints}) -H ~~~,\cr
{\rm time~derivatives}=&\frac{1}{2} {\vec \Psi}^{\dagger}  \cdot (-\vec \sigma \times \partial_0 \vec \Psi)~~~,\cr
{\rm constraints}=&-\Psi_0^{\dagger} K -K^{\dagger}\Psi_0~~~,\cr
H=&-\frac{1}{2}\int d^3x  {\vec \Psi}^{\dagger} \cdot   \vec \nabla \times \vec \Psi ~~~.\cr
\end{align}
From this we read off the canonical momentum $\vec P$ conjugate to $\vec \Psi$,
\begin{align}\label{momentaf}
\vec P=& \frac{1}{2} {\vec \Psi}^{\dagger} \times \vec \sigma~~,~~~ {\vec \Psi}^{\dagger} =i \vec P-\vec P \times \vec \sigma~~~.\cr
\end{align}
Using the canonical bracket definition
\begin{align}\label{brac1}
[\Psi_{i\alpha}(\vec x),P_{j\beta}(\vec y)] =&-\delta_{ij}\delta_{\alpha\beta} \delta^3(\vec x- \vec y)~~~,\cr
\end{align}
with $i,j$ spatial indices and $\alpha,\beta$ spinor indices, we get the further brackets
\begin{align}\label{brac2}
[\Psi_{i\alpha}(\vec x),{\Psi}^{\dagger}_{j\beta}(\vec y)] =&-i (\sigma_j \sigma_i)_{\alpha\beta}\delta^3(\vec x-\vec y)~~~,\cr
[K,K^{\dagger}]=&0~~~,\cr
{[K,H]= 0 \; ,}\; \; & \; {[K^{\dagger}, H]= 0 ~~~.} \cr
\end{align}
The second line of Eq. \eqref{brac2}  shows that the constraints $K$ and $K^{\dagger}$ are  first class in the Dirac terminology, and one can verify that
they serve as generators of the fermionic gauge transformations of $\vec \Psi$ and its adjoint introduced above.
In the free Rarita-Schwinger theory there are no second class constraints (constraints for which the mutual brackets are nonzero).  {The third line of Eq. \eqref{brac2} shows that the Hamiltonian $H$ is also first class so that there are no further constraints.  The Lagrange multipliers $\Psi_{0}$ and $\Psi_{0}^{\dagger}$ are left undetermined by the equations of motion.}

We can now apply the standard formula for counting degrees of freedom \cite{henneaux},
\begin{equation}\label{degfreedom}
{\rm degrees~of~freedom}=\frac{1}{2}(N-2F-S)~~~,
\end{equation}
in which $N$ is the number of real canonical variables, $F$ is the number of real first class constraints, and $S$ is the number
of real second class constraints.  In our case we have $N=3 \times 2 \times 2=12$, $F=2\times 2=4$, and $S=0$, giving 2 for the number
of degrees of freedom for free left-handed Rarita-Schwinger fields.

\section{The gauged Rarita-Schwinger theory}
\label{section_gauged_RS}

To go over to the gauged Rarita-Schwinger theory, one makes the minimal coupling replacements
\begin{equation}\label{minimal}
\vec \nabla \to \vec D~~,~~~\partial_0 \to D_0~~~,
\end{equation}
with $\vec D$ and $D_0$ the space and time components of the four-vector gauge covariant derivative
\begin{equation}\label{minimal1}
D_{\nu}\equiv \partial_{\nu}+ g A_{\nu}~~~,
\end{equation}
where $A_{\nu}$ is the gauge potential, which can be Abelian or non-Abelian.  Apart from this replacement,
the only change in the formulas of the preceding section is in the second line of Eq. \eqref{brac2},
which becomes
\begin{equation} \label{brac3}
[K(\vec x),K^{\dagger}(\vec y)]=-\frac{i}{2} g \vec \sigma \cdot \vec B \delta^3(\vec x-\vec y)~~~,
\end{equation}
with $\vec B$ the magnetic field part of the  gauge field.  Thus $K$ and $K^{\dagger}$ are now second class
constraints, and corresponding to this one finds that the gauged action does not have an off-shell gauge invariance
(although as discussed in \cite{adler2}, it has an on-shell invariance when a secondary constraint following from
the equations of motion is invoked.)   {The Lagrange multipliers $\Psi_{0}$ and $\Psi_{0}^{\dagger}$ are completely determined by the equations of motion \cite{adler2}.}

In the degrees of freedom formula of Eq. \eqref{degfreedom} one now has
$N=12$ as before, but $F=0$ and $S=2\times 2=4$,  giving  4  for the number
of degrees of freedom for gauged left-handed Rarita-Schwinger fields.  Thus, contrary to what was suggested in \cite{adler2}, the
number of degrees of freedom in the gauged case is enlarged relative to the free case.  The discontinuity in the number of degrees of freedom as $g \rightarrow 0$ would not in itself be a problem if the new degrees of freedom behaved properly.  This question is analysed below, after a more manageable reformulation of the gauged theory is recalled.

\section{The extended gauged Rarita-Schwinger theory}
\label{section_extended_RS}

We {thus} turn now to the extended gauged theory introduced in \cite{adler2}, which has an exact off-shell fermionic gauge invariance.
This is achieved by introducing a dimension $\frac{1}{2}$ spin-$\frac{1}{2}$ field $\Lambda$ coupled to the both the gauge fields
and the Rarita-Schwinger field.  Writing the action in the Hamiltonian form of Eq. \eqref{lagrang}, we have
\begin{align}\label{lagrang1}
S=&\int dt L ~~~,\cr
L=&\int d^3x({\rm time~deivatives} + {\rm constraints}) -H ~~~,\cr
{\rm time~deivatives}=&\frac{1}{2} {\vec \Psi}^{\dagger}  \cdot (-\vec \sigma \times \partial_0 \vec \Psi)
-\frac{1}{2}ig \Lambda^{\dagger} \vec \sigma \cdot \vec B \partial_0 \Lambda~~~,\cr
{\rm constraints}=&-\Psi_0^{\dagger} K -K^{\dagger}\Psi_0~~~,\cr
K=&\frac{1}{2} \vec \sigma \cdot \vec D \times \vec \Psi-\frac{1}{2}ig\vec \sigma \cdot \vec B \Lambda~~~,\cr
K^{\dagger}=&-\frac{1}{2} {\vec \Psi}^{\dagger}\cdot (\overleftarrow D \times \vec \sigma)+\frac{1}{2} ig \Lambda^{\dagger} \vec \sigma
\cdot \vec B~~~,\cr
H=&-\frac{1}{2}\int d^3x [  {\vec \Psi}^{\dagger} \cdot   ( \vec D \times \vec \Psi - \vec \sigma \times gA_0\vec \Psi)\cr
-&ig {\vec \Psi}^{\dagger} \cdot \vec C \Lambda+ig \Lambda^{\dagger} \vec C \cdot \vec \Psi
+ig \Lambda^{\dagger} \vec C \cdot \vec D \Lambda  -ig^2 \Lambda^{\dagger} \vec \sigma \cdot \vec B A_0 \Lambda]~~~,\cr
\vec C=&\vec B + \vec \sigma \times \vec E~~~.\cr
\end{align}
As shown in \cite{adler2}, this action is invariant under the gauge transformation
\begin{equation}\label{shiftinv}
\Psi_0 \to \Psi_0+D_0\epsilon~,~~ \vec \Psi \to \vec \Psi+ \vec D \epsilon~,~~\Lambda \to \Lambda -\epsilon~~~.
\end{equation}
From Eq. \eqref{lagrang1} we read off the canonical momenta $\vec P$ and $P$ conjugate respectively to $\vec \Psi$ and $\Lambda$,
\begin{align}\label{momenta1}
\vec P=& \frac{1}{2} {\vec \Psi}^{\dagger} \times \vec \sigma~~,~~~ {\vec \Psi}^{\dagger} =i \vec P-\vec P \times \vec \sigma~~~,\cr
P=& \frac{1}{2}ig \Lambda^{\dagger} \vec \sigma \cdot \vec B~~,~~~\Lambda^{\dagger}=\frac{2}{ig}P (\vec \sigma \cdot \vec B)^{-1}~~~.\cr
\end{align}
Using the canonical bracket definitions
\begin{align}\label{brac4}
[\Psi_{i\alpha}(\vec x),P_{j\beta}(\vec y)] =&-\delta_{ij}\delta_{\alpha\beta} \delta^3(\vec x- \vec y)~~~,\cr
[\Lambda_{\alpha}(\vec x),P_\beta(\vec y)]=&-\delta_{\alpha\beta} \delta^3(\vec x- \vec y)~~~,
\end{align}
with $i,j$ spatial indices and $\alpha,\beta$ spinor indices, we get the further brackets
\begin{align}\label{brac5}
[\Psi_{i\alpha}(\vec x),{\vec \Psi}^{\dagger}_{j\beta}(\vec y)] =&-i (\sigma_j \sigma_i)_{\alpha\beta}\delta^3(\vec x-\vec y)~~~,\cr
[\Lambda_{\alpha}(\vec x), \Lambda^{\dagger}_\beta(\vec y)]=&\frac{2i}{g}\frac{(\vec \sigma)_{\alpha\beta}\cdot \vec B}{\vec B^2} \delta^3(\vec x-\vec y)~~~,\cr
[K,K^{\dagger}]=&0~~~.\cr
\end{align}
The last line of Eq. \eqref{brac4} shows that by virtue of the auxiliary field contributions,
the constraints $K$ and $K^{\dagger}$ in the extended gauged theory have become first class!
Correspondingly, the constraints $K^{\dagger}$ and  $K$  generate the gauge transformation of Eq. \eqref{shiftinv} on $\vec \Psi\,,\Lambda$  (and their adjoints) under the bracket operation of Eq. \eqref{brac4}.  For example, noting that $K= \vec P \cdot \overleftarrow{D} +P$, we have
\begin{align}\label{generation}
[\vec \Psi(\vec x), \int d^3y K^{\dagger}(\vec y) \epsilon(\vec y)]=&{\vec D}_{\vec x}\epsilon(\vec x)~~~,\cr
[\Lambda(\vec x), \int d^3y K^{\dagger}(\vec y) \epsilon(\vec y)]=&-\epsilon(\vec x)~~~.\cr
\end{align}

We can again count degrees of freedom, using the general formula of Eq. \eqref{degfreedom}.  For the Rarita-Schwinger field, we again have
$N=12$, $F=4$, and $S=0$, giving 2 degrees of freedom.  But for the auxiliary field we have $N=2 \times 2=4$, and $F=S=0$, giving 2 additional
degrees of freedom, making 4 in all, in agreement with the counting result for the gauged theory given in Sec. { \ref{section_gauged_RS}}.

Since we are now dealing with an off-shell gauge invariant theory, we can introduce gauge fixing conditions as additional constraints, so that the original first class constraints become second class. We shall follow the convention of labeling constraints involving only $\vec \Psi$ and $\Lambda$ as
$\phi_{1,2}$, and labeling constraints involving only $\vec \Psi^{\dagger}$ and $\Lambda^{\dagger}$, or equivalently the conjugate momenta $\vec P$ and $P$, as
$\chi_{1,2}$. One of the $\phi$ will be proportional to  $K$, and the other  $\phi$ will be a gauge fixing constraint; similarly, one of the $\chi$ will be proportional to $K^{\dagger}$, and the other will be the adjoint gauge fixing constraint.  The nonvanishing brackets of the constraints will be
denoted by
\begin{equation}\label{bracketmatrix}
M_{ab}(\vec x-\vec y)=[\phi_a(\vec x),\chi_b(\vec y)]~~~,
\end{equation}
and in terms of $M$ the Dirac bracket of any $F(\vec \Psi)$ with any  $G(\vec \Psi, \vec \Psi^{\dagger})$ is given by
\begin{equation}\label{dirac}
[F,G]_D=[F,G]-\sum_a\sum_b [F,\chi_a]M_{ab}^{-1}[\phi_b,G]~~~.
\end{equation}
We now proceed to give the results of two specific choices of the gauge fixing constraints.

\section{$\Lambda=0$ gauge}
\label{section_Lambda_0}

We first repeat the bracket calculation in $\Lambda=0$ gauge, to see that this reduces to what is obtained from the
unextended Rarita-Schwinger action.  The constraints now are
\begin{align}\label{lambdazeroconstr}
\phi_1=&\Lambda~~~,\cr
\phi_2=&\vec \sigma \times \vec D \cdot \vec \Psi-ig \vec \sigma \cdot \vec B \Lambda~~~,\cr
\chi_1=&2(\vec P \cdot \overleftarrow D + P)~~~,\cr
\chi_2=&P~~~,\cr
\end{align}
which obey
\begin{equation}\label{adjoint2}
\phi_2^{\dagger}=\chi_1,~~,~~~\phi_1^{\dagger}=\Lambda^{\dagger}=\frac{2}{ig}P(\vec \sigma \cdot \vec B)^{-1}=\frac{2}{ig}\chi_2(\vec \sigma \cdot \vec B)^{-1}~~~.
\end{equation}
For the bracket matrix we find
\begin{align}\label{bracmatrix2}
M_{ab}(\vec x,\vec y)=&[\phi_a(\vec x),\chi_b(\vec y)]=\left(\begin{array}{cc}
-2&-1\\
0&ig\vec \sigma \cdot \vec B\\
\end{array} \right) \delta^3(\vec x-\vec y)~~~,\cr
M^{-1}_{ab}(\vec x,\vec y)=&\left(\begin{array}{cc}
-\frac{1}{2}&-\frac{1}{2ig \vec \sigma \cdot \vec B}\\
0&\frac{1}{ig\vec \sigma \cdot \vec B}\\
\end{array} \right) \delta^3(\vec x-\vec y)~~~.\cr
\end{align}
From Eqs. \eqref{lambdazeroconstr}-\eqref{bracmatrix2}, we find
the following Dirac brackets
\begin{align}\label{lambdazerobracs}
[\Lambda(\vec x),\Lambda^{\dagger}(\vec y)]_D=&[\Lambda(\vec x),\Psi_j^{\dagger}(\vec y)]_D=0~~~,\cr
[\Psi_i(\vec x), \Psi_j^{\dagger}(\vec y)]_D=&-i \sigma_j \sigma_i \delta^3(\vec x-\vec y) + 2i \vec D_{xi}\frac{\delta^3(\vec x-\vec y)}{g\vec \sigma
\cdot \vec B} \overleftarrow D_{yj} \cr
=&
-2i\left[ (\delta_{ij}-\frac{1}{2}\sigma_i\sigma_j)\delta^3(\vec x-\vec y)-\vec D_{xi} \frac{\delta^3(\vec x-\vec y)}{g\vec \sigma
\cdot \vec B} \overleftarrow D_{yj}\right] ~~~.
\end{align}
These agree with the results obtained by first setting $\Lambda=0$ and calculating Dirac brackets in the unextended Rarita-Schwinger theory, in which
the constraints are second class.  This gives a consistency check on the formalism.

\section{Extended gauge covariant radiation gauge}
\label{section_radiation_gauge}

Since the auxiliary field $\Lambda$ has mass dimension $\frac{1}{2}$, rather than the standard $\frac{3}{2}$ of a fermion field, we are
free to add a multiple of $\vec \sigma \cdot \vec B \Lambda$ to $\vec D \cdot \Psi$ to form an extended gauge covariant radiation gauge
constraint.  The choice $0=\phi_1 = \vec D \cdot \vec\Psi-g\vec \sigma \cdot \vec B \Lambda$ leads to particularly simple formulas.
To see that this condition is attainable, we note that under the gauge transformation of Eq. \eqref{shiftinv}, $\phi_1$ transforms as
\begin{equation}\label{phi1trans}
\phi_1 \to \phi_1 + (\vec D^2+g \vec \sigma \cdot \vec B) \epsilon =\phi_1 + (\vec \sigma \cdot \vec D)^2 \epsilon~~~.
\end{equation}
Hence as long as $(\vec \sigma \cdot \vec D)^2$ is invertible, the constraint $\phi_1=0$ is attainable.

Let us define the inverse ${\cal D}$ of $(\vec \sigma \cdot \vec D)^2$ by the equations
\begin{align}\label{inverse}
(\vec \sigma \cdot \vec D_x)^2 {\cal D}  (\vec x-\vec y)=& -\vec \sigma \cdot \vec D_x {\cal D}  (\vec x-\vec y)
\vec \sigma \cdot {\overleftarrow D}_y  ={\cal D}  (\vec x-\vec y)
(\vec \sigma \cdot {\overleftarrow D}_y)^2=\delta^3(\vec x-\vec y)~~~,\cr
{\cal D}(\vec x-\vec y)^{\dagger}=& {\cal D}(\vec y-\vec x)~~~.\cr
\end{align}
Then if initially $\phi_1$ has a nonzero value, it can be shifted to zero by the gauge change of Eq. \eqref{phi1trans} with $\epsilon$ given
by
\begin{equation}\label{gaugeshift}
\epsilon(\vec x) = - \int d^3y {\cal D}(\vec x-\vec y) \phi_1(\vec y)~~~.
\end{equation}

The constraints that we use for gauge covariant radiation gauge are as follows,
\begin{align}\label{constraintset1}
\phi_1=&\vec D \cdot \Psi-g\vec \sigma \cdot \vec B \Lambda~~~,\cr
\phi_2=&\vec \sigma \times \vec D \cdot \vec \Psi-ig \vec \sigma \cdot \vec B \Lambda~~~,\cr
\chi_1=&2(\vec P \cdot \overleftarrow D + P)~~~,\cr
\chi_2=&\vec P \cdot (\vec \sigma \times \overleftarrow D) -i P~~~.\cr
\end{align}
The constraints $\chi_a$ are linear combinations of the adjoints of the constraints $\phi_a$,
\begin{equation}\label{adjoint}
\phi_2^{\dagger}=\chi_1~~,~~~\phi_1^{\dagger}=\frac{1}{2}i\chi_1-\chi_2~~~.
\end{equation}
The nonvanishing brackets of the constraints are given by
\begin{equation}\label{bracketmatrix1}
M_{ab}(\vec x-\vec y)=[\phi_a(\vec x),\chi_b(\vec y)]= 2 ~1_{ab} (\vec \sigma \cdot \vec D_x)^2 \delta^3(\vec x-\vec y)~~~,
\end{equation}
with $1_{ab}$ the $2\times 2$ unit matrix.  So the inverse of the bracket matrix is
\begin{equation}\label{inverse1}
M^{-1}_{ab}(\vec x-\vec y)=\frac{1}{2} 1_{ab} {\cal D}(\vec x-\vec y)~~~.
\end{equation}

We can now compute Dirac brackets using Eq. \eqref{dirac}, with the following results,
\begin{align}\label{diracbrackets}
[\Psi_i(\vec x),\Psi_j^{\dagger}(\vec y)]_D=& -i\sigma_j \sigma_i \delta^3(\vec x-\vec y) -i \vec D_{xi} {\cal D}(\vec x-\vec y)
\overleftarrow D_{yj}\cr
 +&(\vec \sigma \times \vec D_x)_i  {\cal D}(\vec x-\vec y) \overleftarrow D_{yj}
- \vec D_{xi} {\cal D}(\vec x-\vec y) (\vec \sigma \times \overleftarrow D_ y )_j~~~,\cr
[\Psi_i(\vec x),\Lambda^\dagger(\vec y)]_D=&2i (\vec D_x + \frac{1}{2}i \vec \sigma \times \vec D_x)_i {\cal D}(\vec x-\vec y)~~~,\cr
[\Lambda(\vec x),\Psi_j^{\dagger}(\vec y)]_D=& 2i {\cal D}(\vec x-\vec y) ({\overleftarrow D}_y-\frac{1}{2} i \vec \sigma \times {\overleftarrow D}_y)_j~~~,\cr
[\Lambda(\vec x),\Lambda^\dagger(\vec y)]_D=& \frac{2i}{g} \frac{\vec \sigma \cdot \vec B}{\vec B^2} \delta^3(\vec x-\vec y) -
3i {\cal D}(\vec x-\vec y)~~~.\cr
\end{align}

We see from these covariant radiation gauge formulas that the Dirac bracket $[\Psi_i(\vec x),\Psi_j^{\dagger}(\vec y)]_D$ is nonsingular for small $\vec B$; the small $\vec B$ singularity found in \cite{johnson} and \cite{velo} is present only in the auxiliary field bracket $[\Lambda(\vec x),\Lambda^\dagger(\vec y)]_D$.  We also can verify that
\begin{equation}\label{sigmadot}
\sigma_i [\Psi_i(\vec x),\Psi_j^{\dagger}(\vec y)]_D=[\Psi_i(\vec x),\Psi_j^{\dagger}(\vec y)]_D\sigma_j=0~~~.
\end{equation}
This is a direct consequence of the fact that
\begin{equation}\label{choice}
\vec \sigma \cdot \vec D_x  \sigma_i \Psi_i= (\vec D_x+ i \vec \sigma \times \vec D_x)_i \Psi_i=\phi_1+i\phi_2~~~,
\end{equation}
which was the motivation for the specific choice of the extended covariant gauge constraint $\phi_1$.

To study the positivity of Dirac brackets when mapped to anticommutators, we follow the method used in Eqs. (36) and (37) of \cite{adler3}.
Defining (for $F$ either $\Psi_i$ or $\Lambda$)
\begin{equation}\label{tildedef}
\tilde F=F-\sum_{a,b}[F,\chi_a] M^{-1}_{ab}\phi_b~~~,
\end{equation}
we have (for $G$ either $\Psi^{\dagger}_j$ or $\Lambda^{\dagger}$)
\begin{equation}\label{tildebrac}
[F,G]_D=[\tilde F, \tilde G]~~~.
\end{equation}
Writing
\begin{align}\label{tilde1}
\tilde \Psi_i(\vec x)=&\int d^3y [R_{ij}(\vec x, \vec y)\Psi_j(\vec y)+R_i(\vec x,\vec y) \Lambda(\vec y)]~~~,\cr
\tilde \Lambda(\vec x)=& \int d^3y [R(\vec x,\vec y) \Lambda(\vec y)+ \hat R_i(\vec x, \vec y) \Psi_i(\vec y)]~~~\cr
\end{align}
we find
\begin{align}\label{tilde2}
R_{ij}(\vec x,\vec y)=&\delta_{ij}\delta^3(\vec x-\vec y)+ \vec D_{xi}{\cal D}(\vec x-\vec y) \overleftarrow D_{yj}
+\frac{1}{2}(\vec \sigma \times \vec D_x)_i {\cal D}(\vec x-\vec y)(\vec \sigma \times \overleftarrow D_y)_j~~~,\cr
R_i(\vec x,\vec y)=& g (\vec D_x + \frac{1}{2}i\vec \sigma \times \vec D_x)_i {\cal D}(\vec x-\vec y)\vec \sigma \cdot \vec B(\vec y)~~~,\cr
\hat R_i(\vec x,\vec y)=&-{\cal D}(\vec x-\vec y)(\overleftarrow D_y-\frac{1}{2}i\vec \sigma \times \overleftarrow D_y)_i~~~,\cr
R(\vec x,\vec y)=&\delta^3(\vec x-\vec y)-\frac{3}{2}g{\cal D}(\vec x-\vec y)\vec\sigma \cdot \vec B(\vec y)~~~.\cr
\end{align}
One can now verify the following identities
\begin{align}\label{rident}
\sigma_iR_{ij}=R_{ij}\sigma_j=&0~~~,\cr
\sigma_i R_i=\hat R_i \sigma_i=&0~~~.\cr
\end{align}
From Eqs. \eqref{tildebrac}-\eqref{rident}, one now finds the following alternative expressions for the Dirac brackets
\begin{align}\label{altbrac}
[\Psi_i(\vec x),\Psi_j^{\dagger}(\vec y)]_D=&-2i\int d^3w R_{il}(\vec x,\vec w) R_{jl}^{\dagger}(\vec y,\vec w)\cr
+&2ig\int d^3w (\vec D_x+\frac{1}{2}i \vec \sigma \times \vec D_x)_i {\cal D}(\vec x-\vec w)\vec \sigma \cdot \vec B(\vec w){\cal D}(\vec w-\vec y)
(\overleftarrow D_y-\frac{1}{2}i \vec \sigma \times \overleftarrow D_y)_j~~~,\cr
[\Lambda(\vec x),\Lambda^{\dagger}(\vec y)]_D=&\int d^3w R(\vec x, \vec w) \frac{2i}{g} \frac{\vec \sigma \cdot \vec B(\vec w)}{\vec B(\vec w)^2}
R(\vec y,\vec w)^{\dagger}-2i\int d^3w \hat R_i(\vec x,\vec w) \hat R_i^{\dagger}(\vec y, \vec w)~~~,
\end{align}
which by considerable algebra can be verified to agree with the Dirac brackets of Eq. \eqref{diracbrackets}.

When multiplied by $i$ to convert to an anticommutator, the first term in the first line of Eq. \eqref{altbrac} is positive semedefinite
\big(see Eq. (51) of \cite{adler3}).  So overall positivity depends on a comparison of the first and second terms, which in a special case is
undertaken in the next section.
The anticommutator arising from the auxiliary field Dirac bracket on the second line of Eq. \eqref{altbrac} is singular for small $\vec B$
and is not positive semidefinite; this will also be studied further in the next section.

\section{Failure of positivity of the corresponding anticommutators}
\label{section_failure_positivity}

\subsection{$g=0$ Fourier analyis}

The second line of Eq. \eqref{inverse} implies that the first line of Eq. \eqref{altbrac} can be rewritten as
\begin{equation}\label{altbrac2}
-2i\int d^3w R_{il}(\vec x,\vec w) R_{lj}(\vec w,\vec y)
\end{equation}
in which the indices and vector arguments are in natural matrix multiplication order.
Let us now study  Eqs. \eqref{inverse} and {the $[\Psi_i(\vec x),\Psi_j^{\dagger}(\vec y)]_D$
Dirac bracket on the first line of Eq. \eqref{altbrac}} in the limit  $g=0$ of vanishing gauge coupling, where $\vec D=\vec \nabla$.  Fourier transforming according to
\begin{align}\label{fourier}
\delta^3(\vec x-\vec y)=&(2\pi)^{-3}\int d^3 k e^{i\vec k \cdot (\vec x-\vec y)}~~~,\cr
{\cal D}(\vec x-\vec y)=&(2\pi)^{-3}\int d^3 k D[\vec k] e^{i\vec k \cdot (\vec x-\vec y)}~~~,\cr
R_{ij}(\vec x-\vec y)=&(2\pi)^{-3}\int d^3 k R_{ij}[\vec k] e^{i\vec k \cdot (\vec x-\vec y)}~~~,\cr
\end{align}
we have $D[\vec k]=-1/(\vec k)^2$, and
\begin{equation}\label{general}
R_{ij}[\vec k]=R_{ij}[\hat k]=\delta_{ij}-\hat k_i \hat k_j-\frac{1}{2} (\vec \sigma \times \hat k)_i (\vec \sigma \times \hat k)_j~~~,
\end{equation}
with $\hat k=\vec k/|\vec k|$ a unit vector.
From this expression for general $\vec k$ , we can verify that $\sigma_iR_{ij}=0$, and
we also see that $\hat k_i R_{ij}[\hat k]=0$, showing that
$\psi_i=\hat k_i \chi^{\dagger}$, with $\chi$ a general spinor, is a zero eigenvector in Fourier space.

Taking $\hat k = \hat z$ one
get the following expression for $R_{ij}[\hat k]$,
\begin{equation}\label{array}
R_{ij}[\hat z]=\left(
\begin{array}{ccc}
\frac{1}{2}&-\frac{1}{2}i\sigma_3&0\\
\frac{1}{2}i\sigma_3&\frac{1}{2}&0\\
0&0&0\\
\end{array}
\right)~~~,
\end{equation}
From this we find that $R^2=R$, showing again there are zero eigevectors, which can be calculated explicitly by first going to
a representation where $\sigma_3$ is diagonal.

\subsection{Small $\vec B$ non-positivity of $i[\Psi_i(\vec x),\Psi_j^{\dagger}(\vec y)]_D$. }
Let us now expand the Fourier transform {of $[\Psi_i(\vec x),\Psi_j^{\dagger}(\vec y)]_D$ in} Eq. \eqref{altbrac} in powers of $g \vec B$, assuming spatially constant $\vec B$.  Writing
\begin{equation}\label{rexp}
R_{ij}[\vec B, \vec k]= R_{ij}[\vec k] + R_{ij}^{(1)}[\vec B, \vec k]~~~,
\end{equation}
with $R_{ij}[\vec k]$ the zeroth order expression of Eq. \eqref{general} and $R_{ij}^{(1)}[\vec B, \vec k]$ a correction that is first order
in $\vec B$.  Then since $\hat k_iR_{ij}[\hat k]=0$, we have
\begin{equation}\label{vanish}
\hat k_i R_{ij}[\vec B, \vec k]=\hat k_i R_{ij}^{(1)}[\vec B, \vec k] = O(\vec B)~~~.
\end{equation}

Consider now a spatial function $f_i(\vec x)$ constructed as
\begin{equation}\label{spatial}
f_i(\vec x)=\int d^3 x e^{-i \vec k \cdot \vec x} \hat k_i  f(|\vec k|),
\end{equation}
with $f(|\vec k|)$ chosen to make the spatial integral converge.  By Eq. \eqref{vanish}, $f_i(\vec x)$ is a
zero eigenvector of $R_{ij}(\vec x,\vec w)$, and so forming
\begin{equation}\label{vanish1}
\int d^3x \int d^3 y f_i(\vec x) f_j^*(\vec y) [\Psi_i(\vec x),\Psi_j^{\dagger}(\vec y)]_D~~~,
\end{equation}
the contribution of the first line of Eq. \eqref{altbrac} is $O((\vec B)^2)$~~~.
But the contribution of the second line, in Fourier space, is proportional to
\begin{align}& \hat k_i\hat k_j | f(|\vec k|)|^2 |\vec k|^{-4} (\vec k + \frac{1}{2} i \vec \sigma \times \vec k)_i
\vec \sigma \cdot \vec B  (\vec k - \frac{1}{2} i \vec \sigma \times \vec k)_j\cr
=  & |f(|\vec k|)|^2 |\vec k|^{-2 } \vec \sigma \cdot \vec B  ~~~,\cr
\end{align}
which is nonzero and indefinite in sign.  Hence $i[\Psi_i(\vec x),\Psi_j^{\dagger}(\vec y)]_D$ is not positive semidefinite for small $\vec B$.

\subsection{Negative semi-definiteness of the $\vec B$ averaged $i[\Lambda(\vec x),\Lambda^{\dagger}(\vec y)]_D$ }

Consider now the $[\Lambda,\Lambda^{\dagger}]_D$
Dirac bracket on the final line of Eq. \eqref{altbrac}.     Multiplying by $i$  to  get the corresponding anticommutator,
and averaging over the sign of $\vec B$, one gets $3<{\cal D}(\vec x-\vec y)>_{AV}$, since the singular term is odd
in $\vec B$  and drops out of the average.  But $\cal D$ is the inverse of $(\vec \sigma \cdot \vec D)^2$, which is
negative semi-definite since $\vec \sigma \cdot \vec D$ is anti-self-adjoint, and so $\cal D$ is also negative semi-definite.
Thus the averaged anticommutator involving the auxiliary field is negative semidefinite, rather than
positive semi-definite.

\section{Path Integral in Covariant Radiation Gauge}
\label{section_path_integral}

Returning to the constraints of Eq. \eqref{constraintset1},  we give the analog in the extended Rarita-Schwinger theory of the path integral construction
of Sec. 6 of \cite{adler3} .  The functional integral must now include an integration over $\Lambda$ and its conjugate momentum $P$. Integrating over
$\Psi_0$ and $\Psi_0^\dagger$, and using the secondary constraint delta functions together with the primary constraint delta functions, we find
the same simplifications as in Sec. 6 of \cite{adler3}, and so only the four constraints of Eq. \eqref{constraintset1} remain in the functional
integration measure.  We then end up with the following path integral formula (with $P=\frac{1}{2}ig\Lambda^{\dagger}\vec \sigma \cdot \vec B$),
\begin{align}\label{path1}
\langle {\rm out} | S |{\rm in} \rangle \propto & \int \exp\left\{i\left[\int d^4x(\partial_0 \Lambda P + \partial_0 \vec \Psi \cdot
\frac{1}{2}\vec \Psi^{\dagger} \times \vec \sigma)-\int dt H\right]\right\} ~\prod_{t,\vec x} d\mu\big(\vec \Psi,\vec \Psi^{\dagger},\Lambda,
P\big)~~~,\cr
 d\mu\big(\vec \Psi,\vec \Psi^{\dagger},\Lambda,P\big)=&\prod_{a=1,2}\delta(\phi_a)\delta(\chi_a)(\det M_{ab})^{-1}d\vec \Psi d\vec \Psi^{\dagger}
 d\Lambda dP~~~,\cr
  H=&-\frac{1}{2} \int d^3x [\vec \Psi^{\dagger} \cdot \vec D \times \vec \Psi+ ig \Lambda^{\dagger}\vec C \cdot \vec D \Lambda
 -g A_0(\vec \Psi^{\dagger} \cdot \vec \sigma \times \vec \Psi+ig \Lambda^{\dagger} \vec \sigma \cdot \vec B \Lambda)\cr
 -&ig \vec \Psi^{\dagger} \cdot \vec C \Lambda +ig \ \Lambda^{\dagger} \vec C \cdot \vec \Psi]~~~,\cr
 \vec C=&\vec B + \vec \sigma \times \vec E~~~.\cr
\end{align}
In a gauge with $A_0=0$, the formula for $H$ simplifies to
\begin{equation}\label{path2}
 H=-\frac{1}{2} \int d^3x [\vec \Psi^{\dagger} \cdot \vec D \times \vec \Psi+ ig \Lambda^{\dagger}\vec C \cdot \vec D \Lambda
 -ig \vec \Psi^{\dagger} \cdot \vec C \Lambda +ig \ \Lambda^{\dagger} \vec C \cdot \vec \Psi]~~~,
 \end{equation}
which when used in Eq. \eqref{path1}  gives the extension of Eq. (70) of \cite{adler3}.

\section{Discussion}
\label{section_discussion}

We have seen that in the extended Rarita-Schwinger theory, which has a full fermionic off-shell gauge invariance but additional degrees of freedom with respect to the non-interacting theory, the canonical anticommutators that
correspond to the covariant radiation gauge Dirac brackets are not positive semidefinite.  This means that canonical quantization cannot be carried out within a conventional positive semidefinite metric Hilbert space.  This leaves several possibilities:
\begin{enumerate}

\item The theory is not quantizable at all, as suggested in \cite{johnson} and \cite{velo}.

\item The theory can be quantized, but requires use of an indefinite metric Hilbert space, as in Lorentz gauge quantum
electrodynamics.  This possibility is suggested by the fact that it is the canonical brackets associated with the auxiliary field
that cause the breakdown of positivity.

\item The theory can be quantized, but the issue of the  Hilbert space signature can be bypassed by getting
Feynman rules directly from the path integral formulation, and then proceeding to calculation of the gauge anomaly.

\item  Consistency of the theory requires additional couplings to standard dimension $\frac{3}{2}$ spin-$\frac{1}{2}$ fermions.  Such couplings may
play a role \cite{adler4} in generating masses for the Rarita-Schwinger fields in the model of \cite{adler1}, and their effect on the analysis
given here remains to be explored.  Non-minimal couplings, as suggested in \cite{Porrati:2009bs}, might play an interesting role in this respect.

\end{enumerate}

\section*{Acknowledgement}
S.~L.~A. wishes to acknowledge the hospitality of the Aspen Center for Physics and its support by the National
Science Foundation under Grant No. PHYS-1066293. {The work of M.~H. is partially supported by the ERC Advanced Grant ``High-Spin-Grav" and by FNRS-Belgium (convention FRFC PDR T.1025.14 and  convention IISN 4.4503.15).  P.~P. is partially supported by Fondecyt Grant 1140155 and also thanks the Faculty of Mathematics and Physics of Charles University in Prague, Czech Republic, for the kind hospitality during part of the development of this work. The Centro de Estudios Cient\'{\i}ficos (CECs) is funded by the Chilean Government through the Centers of Excellence Base Financing Program of Conicyt.}

\appendix
\section{Summary of identities}
\label{appendix}

We note the following identities \cite{adler2} that are used in the sections above:
\begin{align}\label{ident}
\vec D \times \vec D=&\overleftarrow D \times \overleftarrow D= -ig \vec B~~~,\cr
(\vec \sigma \times \vec D)^2=&2 \vec D^2 + g \vec \sigma \cdot \vec B~~~,\cr
(\vec \sigma \cdot \vec D)^2=& \vec D^2 + g \vec \sigma \cdot \vec B~~~,\cr
\vec D \cdot (\vec \sigma \times \vec D)=&ig \vec \sigma \cdot \vec B~~~,\cr
(\vec \sigma \times \overleftarrow D)\cdot \overleftarrow D =&-ig \vec \sigma \cdot \vec B~~~,\cr
\vec  \sigma \times \vec \sigma =&2i\vec \sigma  ~~~,\cr
\vec \sigma \cdot \vec  v\sigma_j=&v_j+i(\vec \sigma \times \vec v)_j~~~,\cr
\sigma_j \vec \sigma \cdot \vec  v=& v_j-i(\vec \sigma \times \vec v)_j~~~,\cr
(\vec \sigma \times \vec v)_i \sigma_j \sigma_i=&2iv_j~~~.\cr
\end{align}

\end{document}